\documentclass[12pt]{article}

\newcommand{\be}{\begin{equation}}
\newcommand{\ee}{\end{equation}}
\newcommand{\beqs}{\begin{eqnarray}}
\newcommand{\eeqs}{\end{eqnarray}}
\newcommand{\LL}{{\partial_\tau}}
\newcommand{\DD}{{D_\tau}}
\newcommand{\half}{{\frac{1}{2}}}
\newcommand{\tr}{{\rm tr}}

\newcommand{\hV}{{\hat V}}
\newcommand{\thab}{{\theta^{\alpha \beta}}}
\newcommand{\thmn}{{\theta^{\mu \nu}}}
\newcommand{\st}{{\vartheta}}

\newcommand{\thRa}{{\theta_b^{D\alpha}}}

\newcommand{\thoa}{{\theta_{+}^{D\alpha}}}

\newcommand{\thabp}{{\theta_+^{\alpha \beta}}}

\newcommand{\thij}{{\theta^{ij}}}

\newcommand{\thRao}{{\theta_o^{D\alpha}}}

\newcommand{\thabo}{{\theta_o^{\alpha\beta}}}
\newcommand{\thbao}{{\theta_o^{\beta\alpha}}}

\expandafter\ifx\csname mathbbm\endcsname\relax

\else

\fi
\begin{document}
\begin{titlepage}
\begin{flushleft}  
       \hfill                       CCNY-HEP-04/10\\
       \hfill                       November 2004\\
\end{flushleft}
\vspace*{3mm}
\begin{center}
{\LARGE {Kac-Moody theories for colored phase space (quantum Hall) droplets
}\\}
\vspace*{12mm}
\large Alexios P. Polychronakos \\
\vspace*{5mm}
\large
{\em Physics Department, City College of the CUNY\\
Convent Avenue and 138$^{th}$ Street, New York, NY 10031\\
\small alexios@sci.ccny.cuny.edu \/}\\
\vspace*{4mm}
\vspace*{15mm}
\end{center}

\begin{abstract} {
We derive the canonical structure and hamiltonian for arbitrary
deformations of a higher-dimensional (quantum Hall) droplet
of fermions with spin or color on a general phase space manifold.
Gauge fields are introduced via a Kaluza-Klein construction on the
phase space.  The emerging theory is a nonlinear higher-dimensional
generalization of the gauged Kac-Moody algebra.
To leading order in $\hbar$ this reproduces the edge state chiral
Wess-Zumino-Witten action of the droplet.}

\end{abstract}

\end{titlepage}

\section{Introduction}

Describing fermions in terms of bosonic variables has been the source of
much of our progress in understanding their many-body dynamics. 
Such descriptions are collectively termed bosonization \cite{CoMa}-\cite{Pol}.

An intuitive approach to such a description is to consider a dense collection
of fermions forming a `droplet' on their phase space and study the dynamics
of the Fermi surface of the droplet. In two dimensions this leads to a
chiral theory in 1+1 dimensions \cite{Pol}.

In a previous paper \cite{Polcar} we used a phase space canonical
approach to derive the Poisson structure and hamiltonian for arbitrary
deformations of constant-density droplets for spinless (abelian) fermions
on a general higher-dimensional phase space.
This provided a nonlinear generalization of the results of Karabali and
Nair \cite{KarNa} who derived a chiral boundary action for the
higher-dimensional quantum Hall effect proposed by Zhang and Hu
\cite{ZhaHu}. The nonlinear terms derived in \cite{Polcar}
captured higher order quantum corrections in the $1/N$ approximation.

The analysis in \cite{KarNa} uses the quantum density matrix formulation
and large-N approximations \cite{Sak} and also applies to a nonabelian
situation \cite{KaNa}, where it leads to a generalization of the chiral
sigma model known as the Wess-Zumino-Witten model \cite{WZW}. This model
describes boundary perturbations of the quantum Hall (droplet) state of
fermions with color degrees of freedom.

The inclusion of spin, color or other internal degrees of freedom for the
fermions in the phase space formulation presents new challenges,
since the usual semiclassical droplet picture
has to be modified and extended to accommodate the new degrees of freedom.
Nevertheless, such a description is possible and leads to an interesting
gauge generalization of the results for scalar particles. This will be
derived in the present paper.

The organization of the paper is as follows. In section {\bf 2} we give a
general analysis of phase space density dynamics and review the relevant
results of \cite{Polcar}.
In section {\bf 3} we introduce internal degrees of freedom as classical
phase space variables and derive the corresponding droplet dynamics;
we argue that the internal space needs to be quantized for an accurate
description of the system and present the corresponding dynamics in
terms of matrix generalizations of the boundary field obeying a
generalized Kac-Moody algebra. We further demonstrate that the dynamics
to leading
order in $\hbar$ reproduce the Wess-Zumino-Witten model.
In section {\bf 4} we introduce gauge degrees of freedom and define gauge
transformations in the phase space structure, through a mechanism
analogous to a quantum Kaluza-Klein reduction; we generalize the
results of the previous section for the gauged case, reproducing a
gauge Kac-Moody algebra and a gauged Wess-Zumino-Witten model.
Finally, in the last section we discuss some outstanding issues.

{\it Note on $\hbar$:} It is customary to put $\hbar = 1$ and eliminate
it from all expressions. In our case, however, we have to keep track
of various orders of $\hbar$ in our calculations in order to reproduce
the correct leading-order model. We could have kept the convention
$\hbar=1$ and re-introduce it where appropriate and needed. For the
sake of clarity, however, we preferred to keep $\hbar$ explicit
as a book-keeping device, in order to indicate the appropriate scale
of each term in the formulae, accepting the price of some $\hbar$-litter
in the expressions.

\section{Review of phase space droplet dynamics}
\subsection{General formulation}

We shall start by considering noninteracting (spinless) particles on a general
$D$-dimensional phase space manifold with coordinates
$\phi^\mu$, $\mu = 1,\dots D$ and Poisson structure
\be
\{ \phi^\mu , \phi^\mu \}_{sp} = \thmn
\label{spPB}
\ee
where the subscript $sp$ stands for single-particle.
For a non-degenerate Poisson structure the dimension $D$ should be even.
The volume element in this phase space is
\be
d^D \phi = \frac{d\phi}{\sqrt{\theta}} ~~,~{\rm where}~~
d\phi = \prod_{\mu=1}^D d \phi^\mu ~,~~ \theta = \det\thmn
\ee
The particles have hamiltonian $V(\phi)$ and perform classical motion:
\be
{\dot \phi}^\mu = \{ \phi^\mu, V \}_{sp} = \thmn \partial_\nu V
\ee

A dense collection of particles on this phase space is described in terms
of its density $\rho (\phi,t)$. (This is essentially a phase space fluid
dynamical description; for a recent review see \cite{JNPP}.) The motion of the
underlying particles induces a time dependence for the density $\rho$.
Its time evolution is given by a canonical transformation generated by $V$:
\be
{\dot \rho} = \{ \rho , V \}_{sp} = \thab \partial_\alpha \rho
\partial_\beta V
\label{eomrho}
\ee
We can obtain the same dynamics without referring to the underlying
particles by assuming a hamiltonian and canonical structure
for the field $\rho$
\be
{\dot \rho} = \{ \rho , H \}
\ee
[These brackets should not
be confused with the single-particle brackets (\ref{spPB}).] 
Choosing as hamiltonian the total particle energy
\be
H = \int \frac{d\phi}{\sqrt{\theta}} \rho V
\label{H}
\ee
the appropriate Poisson brackets are
\be
\{ \rho(\phi_1 ) , \rho(\phi_2 ) \} = \sqrt{\theta(\phi_+ ) }
\thmn (\phi_+ ) \partial_\mu \rho (\phi_+ ) \partial_\nu \delta
(\phi_- ) 
\label{P}
\ee
where we defined relative and mid-point coordinates $\phi_- = \phi_1
- \phi_2$ and $\phi_+ = \frac{\phi _1 + \phi_2}{2}$. 

The above brackets (\ref{P}) are the standard infinite-dimensional
Poisson algebra of functions on the phase space manifold.
In terms of test functions their form becomes more obvious: defining
\be
\rho[F] = \int \frac{d\phi}{\sqrt{\theta}} F(\phi) \rho(\phi)
\ee
for some function on the phase space $F$, then the brackets of two 
such integrals are
\be
\{ \rho[F] , \rho [G] \} = \rho[ \{ F, G \}_{sp} ]
\label{PB}
\ee
In deriving the above we used the identity
\be
\partial_\mu \left( \frac{\thmn}{\sqrt{\theta}} \right) = 0
\label{ident}
\ee
which is a corollary of the Bianchi identity for $\thmn$.
The equation of motion (\ref{eomrho}) arises as the canonical evolution
with hamiltonian (\ref{H}).

The above algebra has Casimirs. For any function of a single variable $f(x)$,
the integral
\be
C[f] = \int \frac{d\phi}{\sqrt{\theta}} f(\rho) 
\ee
has vanishing Poisson brackets with $\rho$ and constitutes a Casimir. There
are, thus, an infinite tower of Casimirs spanned
by $C_n \equiv C[x^{n+1} ]$ for $n=0,1,2,\dots$.

\subsection{Droplet dynamics}

We now specialize to the case where the particles underlying
the density $\rho$ are (temporarily spinless) fermions.
A dense collection of fermions in phase space will form a 
Fermi liquid. Semiclassically, the fermions will fill densely a region of the
phase space, with one particle per volume $(2\pi\hbar)^{D/2}$, forming
a constant density droplet of arbitrary shape.
Under time evolution, the density remains constant
inside the droplet (by Liouville's theorem) while each point on its
boundary moves according to the single-particle equation of motion, thus
deforming the shape of the droplet.

To describe the droplet it suffices to determine the shape of its
$D-1$-dimensional boundary. This can be done by expressing one of the
phase space coordinates on the boundary, say $\phi^D \equiv R$, as a 
function of the remaining phase space coordinates $\sigma^\alpha$.
(In the sequel we use early greek letters for the set of indices 
$\alpha = 1,2, \dots D-1$ while middle greek letters will take values in
the full $D$-dimensional space.) So the dynamical
variable is the function $R(\sigma,t)$. (For a finite droplet, it is
convenient to assume that the origin of coordinates is inside the droplet
and to think of $R$ as a `radial' coordinate and $\sigma^i$ as `angular'
coordinates.)

The canonical structure of the droplet variable $R$ arises from a hamiltonian
reduction of the full density canonical structure: a constant-density droplet
of arbitrary shape constitutes a particular class of density functions and
thus a submanifold of the full manifold of configurations for $\rho$,
of the form
\be
\rho = \rho_o \st (R-\phi^D )
\ee
where $\st (x) = \half [1+ \rm{sgn} (x)]$ is the step function.
To find the droplet Poisson brackets we need to project the
canonical two-form of $\rho$ on this submanifold.
This can be done with the help of the so-called cartographic transformation
of the density $\rho$. (We refer the reader to \cite{Polcar}
for further details and a full derivation).
Introducing the shorthand $f_b$ for quantities defined
on the boundary of the droplet,
\be
f_b \equiv f(\phi^D = R , \sigma^\alpha )
\ee
the induced Poisson brackets for $R$ are expressed as
\be
\{ R(\sigma_1 ) , R(\sigma_2 ) \} = \frac{\sqrt{\theta_{b+}}}{\rho_o} 
\left[ \theta_{b+}^{D\alpha} \partial_\alpha \delta (\sigma_- ) 
- \theta_{b+}^{\alpha\beta} \partial_\alpha R_+ \partial_\beta 
\delta (\sigma_- ) \right]
\label{PR}
\ee
and the hamiltonian becomes
\be
H = \rho_o \int \frac{d\phi}{\sqrt {\theta}} V(\phi) \st (R-\phi^D ) 
\ee
As before, $+$ and $-$ stand for mid-point and relative $\sigma$ coordinates.

The above Poisson structure and hamiltonian encode the full dynamics
of the droplet and imply the canonical evolution for the boundary
\be
{\dot R} = \thRa \,\partial_\alpha V_b -\theta_b^{\alpha \beta} 
\,\partial_\alpha R \,\partial_\beta V_b
\label{eomR}
\ee
This equation refers only to the boundary,
although the hamiltonian is defined in the bulk of the droplet.
The same equation can be obtained by following the single-particle
evolution of the particles on the boundary of the droplet \cite{Polcar}.

It is worth noting
that if the coordinate $\phi^D$ is chosen to parametrize the potential
(that is, surfaces $\phi^D$=constant are equipotential), $V=V(\phi^D )$,
the second term above drops and we get
\be
{\dot R} = \thRa \partial_\alpha V_b
\ee
In the special case when $\thRa$ is nonzero only for a single value of the
index $\alpha$ (there is a global variable conjugate to $\phi^D$, call it 
$\phi^1$), and is a function only of $\phi^D$, the above equation becomes
\be
{\dot R} = \theta_b^{01} {V'}_b \partial_1 R
\ee
which can easily be solved by a hodographic transformation, interchanging
$R$ and $\phi^1$. 

We should warn that the droplet may have more than one boundaries, depending
on its topology. In such cases we need to introduce several commuting
boundary fields $R_n$, one for each boundary. Similarly, the boundary could
intersect $\sigma$=constant lines at more than one $\phi^D$, in which
case we need again to introduce several boundary fields, one
for each branch, with appropriate matching conditions tying them into
a unique boundary.

We conclude this section with the following remarks:

1. The Poisson brackets (\ref{PR}) of $R$ contain an affine `chiral' part 
(the first term in the bracket) as well as an ordinary Poisson density
structure over the gauge manifold $\{ \sigma^\alpha \}/\phi^D$.
The quotient arises because $\thab$ is degenerate,
being odd-dimensional, and effectively the variable conjugate to
$\phi^D$ drops out.

2. ({\ref{PR}) satisfies the Bianchi identity, as a corollary of the
Bianchi identity of the Poisson brackets for $\rho$, although its 
direct check is highly nontrivial. In the special case when $\thmn$
is independent of $\phi^D$ the affine and linear terms decouple
and individually satisfy the Bianchi identity. In the
generic case, however, both terms are needed to satisfy the identity.
This will be relevant to the case of particles with internal degrees
of freedom.

3. The Casimirs of the original density for the droplet become
$C_n = C_0$. So they are all neutralized,
the only remaining Casimir being the total particle number $C_0 = N$.

4. The constant $\rho_o$ appears both in the hamiltonian and the Poisson
brackets and is irrelevant for classical dynamics.
The semiclassical interpretation of the droplet, however, fixes the value
$\rho_o = 1/(2\pi \hbar)^{D/2}$, which will be important for quantization
and for the case of spinning particles.

\section{Particles with internal degrees of freedom}

The generalization of the above semiclassical construction for fermions
with internal degrees of freedom (spin, color, flavor etc.) presents some 
conceptual problems, due to the fact that internal quantum numbers are
never really classical.

One approach would be to view the internal degrees of freedom
simply as different species of fermions and apply the above procedure
separately for each species. We would obtain a set of partially
overlapping droplets with mutually commuting boundary fields $R$.
This, however, has several drawbacks. One is that such a description
would not allow the particle number of each species to fluctuate (remember
that the total particle number for each droplet is a Casimir), thus
excluding the possibility of having transitions of fermions from one
species to the other. Another related problem is that the hamiltonian
may not be diagonal in the particular chosen basis of flavors.
This would happen, e.g., for a spin-dependent hamiltonian or in the case
where we include gauge fields that act on the spins or colors.

We need, therefore, to start from a proper semiclassical description
of the full set of degrees of freedom without the above limitations.
This will be done in this section.

\subsection{Introducing spins as classical phase space variables}

We shall start from a description where the classical phase space encodes
also the internal degrees of freedom of the particles. This will be done
by considering the internal quantum numbers as arising from the
quantization of an internal, {\it compact} phase space for the particles.
For shortness, we shall refer to this space as representing
`spin' variables, understanding that it can also represent color, flavor
or any other internal degrees of freedom.

Consider the direct product of the original phase space $\thab$ and an
additional compact phase space with coordinates $\xi^i$ and Poisson
structure $\thij$ (middle latin letters will stand for indices of
the components of this compact phase space). Clearly $\theta^{\alpha i}
= 0$. The dimensionality $D_I$ of the $\xi$ space will be left arbitrary.
The total volume of this space, however, will be chosen as
$n (2\pi\hbar)^{D_I /2}$. We see that the size of this space is microscopic,
involving Planck's constant.

Semiclassically, we will have one quantum state per volume 
$(2\pi\hbar)^{D_I /2}$ in the internal phase space. The above choice of
volume implies that we shall have $n$ quantum states associated
with this space, thus endowing the particles with $n$ internal states.
The classical variables $\xi^i$ represent spin operators for the particle.
The droplet procedure of the
previous section can, then, be applied to the total phase space $(\phi,\xi)$.

We shall choose conventions in which the canonical two-form of the spin
space $\omega_{ij}$ is of order $\hbar$ while the range of the coordinates
$\xi^i$ is of order $\hbar^0$. This makes the Poisson structure 
$\theta^{ij}$ of order $1/\hbar$. We shall also take the internal phase space
to be homogeneous and the determinant of $\thij$ equal to $(2\pi\hbar)^{-D_I}$,
therefore making the determinant of the total Poisson structure
independent of $\xi^i$
\be
\det (\thmn \times \thij) = \det (\thmn ) \det (\thij ) =
(2\pi\hbar)^{-D_I} \theta
\label{dett}
\ee

There are many potential realizations of the spin space $\xi^i$.
A specific example of such a space is $S^2$ with 
canonical two-form proportional to the area form. Choosing $\xi^1 = 
\varphi/2\pi$, $\xi^2 = (n/2) \cos \theta$, with
$(\theta,\varphi)$ polar and azimuthal angles on the sphere, the canonical
two-form and Poisson structure will be
\be
\omega = \frac{\hbar n}{2} \sin\theta \, d\theta d\phi = 2\pi\hbar \,
d\xi^1 d\xi^2 ~\to~ \{ \xi^2 , \xi^1 \} = \frac{1}{2\pi\hbar}
\ee
The range of ($\xi_1 , \xi_2$) is ($1,n$) and the total area of this space
is $2\pi\hbar n$. Semiclassically
it can support $n$ quantum states. The quantization of this phase space 
reproduces the lowest Landau level of a particle on the sphere with
a magnetic monopole of strength $n$ at the center. It is known that these
states form a spin-$\frac{n}{2}$ multiplet of the group of rotations, the
cartesian coordinates of the particle becoming spin-$\frac{n}{2}$ $SU(2)$
matrices. The number of states is $2j+1 = n+1$, the shift due to the nonzero
curvature of the space. 

Another realization of the internal phase space would be 
the grassmanian manifold
$G = U(M)/U(M_1) \times \cdots \times U(M_k)$ ($M_1 + \cdots +M_k = M$)
with an appropriate canonical form. This can be visualized as 
the lowest Landau level of a particle moving on the group manifold $U(M)$
with an appropriate magnetic field. The canonical
structure $\omega = d{\cal A}$ is determined by the Kirillov one-form 
\be
{\cal A} = i \hbar \, \tr ( K U^{-1} d U)
\label{Kir}
\ee
where $U$ is a $U(M)$ matrix and $K$ is a hermitian $M \times M$ matrix
that can be chosen diagonal. The above is invariant under
right-multiplication of $U$ by a unitary matrix commuting with $K$, so 
the corresponding $U(N)$ coordinates have to be eliminated. The phase space
manifold, then, is the grassmanian $G$ where $M_1 , \dots M_k$ are
the degeneracies of the eigenvalues of $K$.

Quantization of the above phase space requires that the eigenvalues
of $K$ be integers (a condition akin to the monopole quantization
on the sphere). The quantum states will form irreducible representations
of $U(M)$ with lengths of Young tableau rows given by the eigenvalues of
$K$ and $U(1)$ charge equal to the total number of boxes. So this phase
space will reproduce internal `color' quantum numbers in a given
representation of $SU(M)$ and the classical coordinates $\xi^i$ represent
color matrices in this representation.
The previous $S^2$ monopole construction can be realized as
the grassmanian manifold $U(2)/U(1)\times U(1)$ with the two eigenvalues
of $K$ differing by $n$.

The exact realization of the spin space is unimportant at this point.
The only thing that matters is the fact that we will have $n$ internal 
states for each fermion. Specific realizations, however, will be more
convenient depending on the dynamics and symmetries of the problem,
as will be apparent later.

\subsection{Realization of droplets with internal degrees of freedom}

We are now set to apply the droplet formalism to the problem. The total
phase space has dimension $D+D_I$ and can accommodate one fermion per
volume $(2\pi\hbar)^{(D+D_I )/2}$. Fermions on this space will form
a droplet with density ${\bar \rho}_o = 1/(2\pi\hbar)^{(D+D_I )/2}$,
reserving the notation $\rho_o = 1/(2\pi\hbar)^{D/2}$ for the density
in coordinate phase space. The droplet boundary
variable $R$ will be a function of both $\sigma^\alpha$ and $\xi^i$.

The Poisson brackets of $R$ are given by the general formula (\ref{PR})
applied to the present phase space structure:
\beqs
\{ R(\sigma_1 , \xi_1 ) , R(\sigma_2 , \xi_2 ) \} =
\frac{\sqrt{\theta_{b+}}}{\rho_o} 
&\Bigl[& 
\theta_{b+}^{D\alpha} \, \partial_\alpha \delta (\sigma_- ) \, \delta (\xi_- )
- \theta_{b+}^{\alpha\beta} \, \partial_\alpha R_+ \, \partial_\beta 
\delta (\sigma_- ) \, \delta (\xi_- ) \cr
&-& \theta_+^{ij} \, \partial_i R_+ \, 
\partial_j \delta (\xi_- ) ~ \Bigr]
\label{PRCS}
\eeqs
where we used (\ref{dett}) and $(2\pi\hbar)^{D_I /2}{\bar \rho}_o = \rho_o$. 
Assuming a particle hamiltonian $V(\phi,\xi)$ that depends also on the
spin variables, the hamiltonian for $R$ will be
\be
H = \rho_o \int \frac{d\phi d\xi}{\sqrt \theta}
\, V(\phi,\xi) \, \st (R(\sigma,\xi) - \phi^D )
\label{Hclasp}
\ee

The above would be an exact classical description of the droplet.
The fact, however, that the internal space is of Planck size and supports
a few quantum states renders it essentially quantum mechanical and makes
the classical description inadequate. We need, therefore, to quantize the
internal degrees of freedom and incorporate them in the Poisson brackets
for $R$. This can be done by quantizing the spin coordinates $\xi$.

This quantization goes along standard lines. The $\xi^i$ become noncommutative
and are represented by $n \times n$ matrices. Functions of the $\xi^i$
become matrices on the $n$-dimensional internal Hilbert space; real functions
such as $R(\xi)$ become hermitian matrices $R^{ab}$, where early latin letters
$a,b,\dots = 1,\dots n$ will stand for spin indices.
Integration over the phase space $\xi$ amounts to summing over Hilbert
space states with a volume of $(2\pi\hbar)^{D_I /2}$ per state, that is,
a trace over the Hilbert space
\be
\int \frac{d\xi}{\sqrt{\det(\thij)}} \to (2\pi\hbar)^{D_I /2} \, \tr
~~~{\rm or}~~~ \int d\xi \to  \tr
\ee
while the $\xi$-Poisson brackets become matrix commutators over $i\hbar$:
\be
\{ A , B \}_\xi \equiv \thij \, \partial_i A \, \partial_j B \to
\frac{1}{i\hbar} [A,B]
\ee
We also need the Dirac delta-function $\delta (\xi_1 - \xi_2 )$. Since
this relates two different points in the $\xi$ space it should carry
two sets of matrix indices $(a_1 , b_1 ; a_2 , b_2 )$ 
and implement the defining property
\be
\int d\xi_1 F(\xi_1 ) \, \delta (\xi_1 - \xi_2 ) = F(\xi_2 )
\ee
This implies for matrix quantities
\be
\tr_1 ( F_1 \delta_{12} ) = F_2
\ee
where the subscripts $1,2$ refer to the first and second set of matrix
indices. This means that $\delta_{12}$ is proportional to the matrix
that exchanges the matrix spaces $1$ and $2$, which is written in terms
of Kronecker deltas as:
\be
(\delta_{12})^{a_1 b_1 ; a_2 b_2} = 
\delta^{a_1 b_2} \, \delta^{a_2 b_1}
\ee
and satisfies
\be
F_1 \delta_{12} = \delta_{12} F_2  ~,~~~
F_2 \delta_{12} = \delta_{12} F_1
\label{exch}
\ee
Now we may determine the dynamics of the matrix field variable $R$.

\subsubsection{Poisson structure} 
In order to obtain Poisson brackets for the matrix variable $R$ in a form
that is not
too unwieldy we shall assume that $\thmn$ is independent of $\phi^D$. This
can always be achieved with an appropriate choice of coordinates;
full generality can be restored after obtaining the
Poisson brackets of $R$ by performing the inverse change of variables.
This choice guarantees that $\theta_b^{\mu \nu}$ and $\sqrt{\theta_b}$
become independent of $R$ and are scalar functions of $\sigma$.

To translate the expression in (\ref{PRCS})
to matrix spin variables, observe that the last term in the Poisson
brackets of $R$, involving $\xi$-space derivatives, can be written as
\be
\theta_+^{ij} \partial_i R_+ \partial_j \delta (\xi_- ) = 
\{ R(\xi_1 ) , \delta (\xi_1 - \xi_2 ) \}_{\xi_1} =
- \{ R(\xi_2 ) , \delta (\xi_1 - \xi_2 ) \}_{\xi_2}
\ee
This translates to the matrix expression
\be
\frac{1}{i\hbar} [ R_1 , \delta_{12} ] =
- \frac{1}{i\hbar} [ R_2 , \delta_{12} ]
\ee
the equality of the two expressions being ensured by (\ref{exch}).
We also have to express functions $F_+$ defined on mid-point coordinates
$\xi_+$ in terms of matrix
indices. For this, we remark that, when $F_+$ multiplies delta functions
or their derivatives, it can also be expressed as
\be
F_+ = F\left( \frac{\xi_1 + \xi_2}{2} \right) = \frac{F(\xi_1 ) + F(\xi_2 )}{2}
\ee
and this translates to $\half (F_1 + F_2 )$ in matrix notation.

We now have all the ingredients. Each classical expression in (\ref{PRCS})
becomes a corresponding matrix in the spin space with a residual
dependence on $\sigma^\alpha$; subscripts $+$ and $-$ refer to mid-point
and relative $\sigma$ coordinates. Altogether we obtain
\beqs
\{ R^{ab} (\sigma_1 ) , R^{cd} (\sigma_2 ) \} =
\frac{\sqrt{\theta_+}}{\rho_o} &\Bigl[& 
\thoa \partial_\alpha \delta (\sigma_- ) \, \delta^{ad} \delta^{cb} \cr
&-& \half \thabp \left( \partial_\alpha R_+^{ad} \delta^{cb} 
+ \partial_\alpha R_+^{cb} \delta^{ad} \right)
\partial_\beta \delta(\sigma_- ) \cr
&-& \frac{1}{i\hbar} \left( R_+^{ad} \delta^{cb} 
- R_+^{cb} \delta^{ad} \right) \delta (\sigma_- ) ~ \Bigr]
\label{Rab}
\eeqs
Note that the last term involves an explicit $\hbar$.

These brackets are a generalization of the Kac-Moody
algebra. To make this explicit, define the generators $T^A$, $A=1,2,\dots
n^2-1$ of the fundamental representation of $SU(n)$ (capital latin letters
will stand for $U(N)$ generator indices) and append the $U(1)$ generator
(proportional to the $n\times n$ unit matrix) $T^0 = I/\sqrt{n}$ for the
full set of generators of $U(n)$. We choose their normalization so
that they satisfy the orthogonality and completeness conditions
\be
\tr ( T^A T^B ) = \delta^{AB} ~,~~~
\sum_{A=0}^{n^2 -1} (T^A)^{ab} (T^A)^{cd} = \delta^{ad} \delta^{cb}
\ee
Their commutators define the $SU(N)$ structure constants
\be
[ T^A , T^B ] = i f^{ABC} T^C
\ee
(with $f^{0AB} = 0$). We further define the symmetrized trace
\be
d^{ABC} = \half \tr (T^A T^B T^C + T^B T^A T^C )
\ee
The symbol $d^{ABC}$ is also known as the anomaly of the group $U(n)$,
since it appears in the evaluation of the triangle anomaly graphs in
$3+1$-dimensional gauge theories.

The matrices $T^A$ can be used as a basis to express the hermitian 
matrix $R$:
\be
R^{ab} = \sum_A R^A (T^A)^{ab} ~,~~~
R^A = \tr (R T^A)
\ee
In terms of the dynamical variables $R^A$ the Poisson brackets (\ref{Rab})
become
\beqs
\{ R^A (\sigma_1 ) , R^B (\sigma_2 ) \} =
\frac{\sqrt{\theta_+}}{\rho_o} &\Bigl[&
\thoa \partial_\alpha \delta (\sigma_- ) \, \delta^{AB} \cr
&-& d^{ABC} \thabp \,\partial_\alpha R_+^C \,\partial_\beta \delta(\sigma_- )
+ \frac{1}{\hbar} f^{ABC} R_+^C \, \delta (\sigma_- ) ~ \Bigr]
\label{RAB}
\eeqs
We recognize this as a Kac-Moody type algebra generalized to higher
dimension with an additional term proportional to the anomaly of $U(n)$.

This algebra inherits the Casimir $C_0$ of the
classical algebra. The integral
\be
N = \rho_o \int \frac{d\sigma}{\sqrt\theta} \, \tr R = 
\rho_o \int \frac{d\sigma}{\sqrt\theta} {\sqrt n} R^0 
\label{Ncasimir}
\ee
commutes with $R$. This is the total particle number.

\subsubsection{Hamiltonian}

The hamiltonian of the droplet can similarly be expressed in matrix form.
The only question is matrix ordering. A $\xi$-dependent single-particle
hamiltonian $V(\phi,\xi)$ and its product with the step function
$\st (R-\phi^D )$ of a matrix $R$ introduce ordering ambiguities.

Ordering will not be an issue as long as $V$ is at most linear in the $\xi^i$.
This is, in fact, related to the question of the specific realization of
the internal spin space, as commented at the end of the previous section
and as we shall analyze presently.

Remember that the spin space is microscopic (of order $\hbar$). The
relevant variables are really $S^i = \hbar \xi^i$, representing quantum
spin operators, and their products are higher order in $\hbar$.
Such higher-order terms will be equally relevant only if they come with large
coefficients (of order $1/\hbar$ for quadratic terms etc.), which is
unnatural. This, however, can be avoided if the internal phase space
is chosen wisely so that all spin terms in the hamiltonian can be
expressed {\it linearly} in the coordinates $\xi^i$. This can best be
illustrated with the following example.

Consider the case that there are $n$ internal states. As exposed in the
previous section, we may realize them either as a spin-$\frac{n-1}{2}$
representation of $SU(2)$, in terms of an $S^2$ phase space, or as the
fundamental representation of $SU(n)$, in terms of the grassmanian space
$U(n)/U(n-1) \times U(1)$, that is, the Kirillov action (\ref{Kir})
with $K = diag (1,0,\dots,0)$.

In either case the Hilbert space is
$n$-dimensional, so there are $n^2$ linearly independent hermitian
operators on this space (including the identity operator) 
that can be used to expand any operator. In
the $SU(2)$ case these are the three spin matrices $S^{1,2,3} = \hbar
\xi^{1,2,3}$ (the $\xi^i$ are Pauli or higher-spin $SU(2)$ matrices)
as well as
their ordered products up to degree $n-1$. A general spin-dependent term in
the single-particle hamiltonian, then, can be expressed as a sum of
such monomials. In the $SU(n)$ realization, however, the $n^2-1$ fundamental
generators (represented by coordinates $\xi^i$) are a complete set and
therefore any hamiltonian can be expressed as a {\it linear} expression
in the $\xi^i$. Clearly, physics originating from $SU(2)$ spin will
only involve linear expressions in the $S^i$, while physics originating
from color or flavor degrees of freedom will give rise to a linear
expression in the full set of $SU(n)$ generators.

In conclusion, an appropriate choice of realization of the internal states
will lead to an expression for $V(\phi,\xi)$ linear in $\xi$ and we may write
\be
V(\phi,\chi) = {\bar V} (\phi) + V_i (\phi) \xi^i \equiv V_0 + \hbar \hV
\ee
where $\bar V$ is the spinless part and $\hbar \hV = V_i \xi^i$ is the spin
part of the single-particle energy (where we explicitly indicate the fact
that it is of order $\hbar$). The $V_i$ are `magnetic field' terms.
The matrix representation of $V$ is unambiguous. Further, there are no
ordering problems in the definition of $\st (R-\phi^D )$, since it is
defined pointwise in $\sigma$ and does not couple matrices
$R$ at different points that could be noncommuting. We may
define it, for instance, via the expression
\be
\st (R-\phi^D ) = -\frac{1}{2\pi i} \int \frac{dk}{k+i\varepsilon}
e^{ik(R-\phi^D )}
\ee
The multiplication of $V$ and $\st (R-\phi^D )$ inside the integral is
also free of ordering issues: $V$ is linear in $\xi$ and the trace
representing $\xi$-integration makes its exact placement immaterial.

Altogether the droplet hamiltonian will be given by (\ref{Hclasp})
with the integral expressed as a trace
\be
H = \rho_o \, \tr \int \frac{d\phi}{\sqrt \theta} \,
V(\phi,\xi) \, \st (R(\sigma,\xi) - \phi^D )
\label{Hsp}
\ee
The matrix expression inside the trace can also be obtained by integrating
the classical expression in terms of $\phi^D$ and then promoting $R$ to a
matrix, which will give the same result as using the matrix
definition of $\st (R-\phi^D )$.

The above Poisson structure and hamiltonian imply an equation of motion
for $R$: ${\dot R} = \{ R,H \}$. In evaluating this Poisson bracket it is
useful to use the relations
\be
F_+ = \frac{F(\sigma_1 ) + F(\sigma_2 )}{2} ~,~~~ 
F_+ G_+ = \frac{F(\sigma_1 ) G(\sigma_2 ) + F(\sigma_2 ) G(\sigma_1 )}{2}
\ee
which hold true for any two functions $F,G$ when multiplying the
delta-function $\delta (\sigma_- )$ or its derivatives. We obtain
\be
{\dot R} = \thRa \,\partial_\alpha V_b -\half \theta_b^{\alpha \beta} 
( \partial_\alpha R \,\partial_\beta V_b + \partial_\beta V_b \,
\partial_\alpha R ) -\frac{1}{i\hbar} [ R,V_b ]
\label{eomatr}
\ee
This is the matrix version of the equation of motion (\ref{eomR}) with
a symmetric ordering of the middle term. 

\subsection{The leading-order action}

Assuming that the droplet deviates slightly from an equilibrium
configuration, we can analyze its motion as a perturbation
of its equilibrium shape. This will be useful for large droplets
(large number of particles) where we can recover a boundary action
to leading order in $\hbar$.

Consider a reference droplet configuration filling the phase space
up to an energy level $E_o$ for the scalar part of the single-particle
hamiltonian $V_0$.
That is, the boundary field of the reference configuration $R=r_o (\sigma)$
is $\xi$-independent (proportional to the identity matrix) and satisfies
\be
V_0 (r_o (\sigma) ,\sigma) = E_o = {\rm constant}
\ee
(From now on, a subscript $_o$ in any quantity will signify the value of
the quantity at $\phi^D = r_o$.)

Note that $r_o (\sigma)$ is not a time-independent solution of the equations
of motion, since it does not minimize the full hamiltonian. 
The true static solution $R_o$ includes a
$\xi$-dependent part $\hbar \chi_o$ and satisfies
\be
V (R_o (\sigma) , \sigma,) = E_o
\ee
Substituting $R_o = r_o + \hbar \chi_o$ and $V = V_0 + \hbar \hV$
in the above and expanding to first order in $\hbar$ we obtain
\be
\chi_o = -\frac{\hV}{u_o} ~~,~~~ {\rm where} ~~~
u_o = \left( \frac{\partial V_0}{\partial \phi^D} \right)_o
\ee
Nevertheless, we may expand our droplet around the reference configuration
$r_o$. Such a perturbation
can be written as 
\be
R = r_o + \hbar \chi
\ee
where $\chi(\sigma)$
is a matrix and we have explicitly indicated that this is an order $\hbar$
perturbation. Correspondingly, the Poisson brackets (\ref{Rab}) or (\ref{RAB})
and hamiltonian (\ref{Hsp}) have to be expanded to that order.

\subsubsection{Leading-order Poisson brackets and hamiltonian}

For the Poisson brackets, the leading part of the first two terms in
(\ref{RAB}) (affine and proportional to $d^{ABC}$)
is of order $\hbar^0$ and involves the
scalar equilibrium term $r_o$ alone. In the last term the scalar
part $r_o$ (proportional to the identity matrix $\delta^{A0}$) drops out;
due to the explicit $1/\hbar$ in its coefficient, however, the $\chi$ part
survives. Overall we have
\be
\{ \chi^A (\sigma_1 ) , \chi^B (\sigma_2 ) \} =
\frac{\sqrt{\theta_{o+}}}{\hbar^2\rho_o} \Bigl[
(\theta_{o+}^{D\alpha} - \theta_{o+}^{\beta\alpha} \,\partial_\beta r_o )
\partial_\alpha \delta(\sigma_- ) \, \delta^{AB}
+ f^{ABC} \chi^C \, \delta (\sigma_- ) \Bigr]
\label{chiAB}
\ee
(Note that we reinstated the dependence of $\thab$ on $R$;
the dependence on the matrix part $\chi$ that would complicate the
Poisson brackets is down by $\hbar$ and can be neglected.)

The single-particle hamiltonian perturbed to order $\hbar$ around
$r_o$ is
\be
V = E_o + \hbar u_o (\sigma) \chi (\sigma) + \hbar \hV_o
\ee
and this gives for the droplet hamiltonian
\be
H = H_o + \hbar \rho_o E_o \int \frac{d\sigma}{\sqrt{\theta_o}} \,\tr \chi +
\hbar^2 \rho_o \int \frac{d\sigma}{\sqrt{\theta_o}} \, \tr \left(
\half u_o \chi^2 + \hV_o \chi \right)
\ee
$H_o$ is the energy of the unperturbed droplet; it is a constant and can
be discarded. The next term, of order $\hbar$, is proportional 
to $E_o (N - N_o )$, where $N$ is the Casimir (\ref{Ncasimir}) and $N_o$
the value of the Casimir for the unperturbed droplet.
It is therefore itself a Casimir and does not contribute to the equations
of motion. It can be set to zero as an initial condition, 
corresponding to a constant-volume perturbation of the droplet
(total number of particles $N$ constant). We end up with a hamiltonian
of order $\hbar^2$ with a linear and a quadratic term in $\chi$:
\be
H = \hbar^2 \rho_o \int \frac{d\sigma}{\sqrt{\theta_o}} \, \tr \left(
\half u_o \chi^2 + \hV_o \chi \right) 
\ee

Finally, the equation of motion as obtained by the above hamiltonian
and Poisson brackets, or from the full equation of motion (\ref{eomatr})
to first order in $\hbar$, becomes
\be
{\dot \chi} = (\thRao -\thbao \partial_\beta r_o )
\partial_\alpha (u_o \chi + \hV_o ) + i [ \chi , \hV_o ]
\label{eomchi}
\ee
Note that the factors $\hbar^2 \rho_o$ have cancelled.

\subsubsection{Wess-Zumino-Witten action}
The form of the above equation of motion is suggestive. Let us define
the differential operator
\be
\LL = ( \thRao - \thbao \partial_\beta r_o ) \partial_\alpha
\ee
representing a vector field along the classical trajectory of a particle
on the boundary of the droplet. Note that $\LL$ is properly anti-self-adjoint
on the boundary of the droplet under the
integration measure $d\sigma/\sqrt{\theta_o}$ due to the relation
\be
\partial_\alpha \left( \frac{\theta^{\mu \alpha}}{\sqrt{\theta_o}} \right) = 0
\ee
In terms of $\LL$ the equation of motion becomes
\be
{\dot \chi} + i [ \hV_o , \chi ]- \LL ( u_o \chi) = \LL \hV_o
\ee
Similarly, the Poisson brackets of $\chi$ become
\be
\{ \chi^A (\sigma_1 ) , \chi^B (\sigma_2 ) \} =
\frac{\sqrt{\theta_{o+}}}{\hbar^2\rho_o} \Bigl[
\LL_+ \delta(\sigma_- ) \delta^{AB}
+ f^{ABC} \chi^C \, \delta (\sigma_- ) \Bigr]
\label{chiKM}
\ee
We recognize the above Poisson structure as the Kac-Moody algebra
of a chiral current $J_+ = \chi$. This algebra is realized by one
of the light-cone components of the current in a Wess-Zumino-Witten (WZW)
model, with the corresponding light-cone coordinate $x^+$ identified as the
trajectory along which $\LL$ acts and the other light-cone coordinate
$x^-$ identified as time. The remaining directions of $\sigma$ appear
simply as parameters.

This immediately  provides a lagrangian realization of the above hamiltonian
structure, in terms of the WZW action of a unitary $n \times n$
matrix field $U$. The chiral field $\chi$ is identified as the current
\be
\chi = -i U^{-1} \LL U
\ee
Interestingly, this form for $\chi$ implies that the deformation of
the droplet is generated by the unitary field $U$, in analogy with the
spinless case. Specifically, consider a small deformation of the coordinates
$\phi,\xi$ generated by an order-$\hbar$ generating function
$\hbar \Phi (\sigma,\xi)$:
\be
r_o \to r_o + \hbar \thRao \partial_\alpha \Phi ~,~~~
\sigma^\alpha \to \sigma^\alpha + \hbar \thabo \partial_\beta \Phi
~,~~~ \xi^i \to \xi^i + \hbar \thij \partial_j \Phi
\ee
The deformations of $\phi^D = r_o$ and
$\sigma^\alpha$ are infinitesimal.
The deformation of $\xi^i$, however, is of order $\hbar^0$ and 
of the same order as $\xi^i$; it cannot be written in the above
infinitesimal form. Instead, we must write the analog of a finite canonical
transformation on the spin space, which is a unitary transformation.

The action that reproduces the Poisson brackets of $\chi$ is
the WZW action at critical coupling. The WZW model gives equation of motion
$\partial_- J_+ = 0$, corresponding to ${\dot \chi} =0$, which implies
a vanishing hamiltonian. The full action, then, is the WZW action minus
the hamiltonian
\be
S = S_{WZ} + \hbar^2 \rho_o \int dt \frac{d\sigma}{\sqrt{\theta_o}} 
\, \tr \left[ -\half
U^{-1} \partial_t U  U^{-1} \LL U + \half u_o (U^{-1} \LL U)^2
+ i \hV_o U^{-1} \LL U \right]
\label{S}
\ee
The Wess-Zumino action $S_{WZ}$ can be written as an integral
over a $D$-dimensional manifold whose boundary is the boundary of the
droplet. Introducing a variable $s \in [0,1]$ we can take
$U(s,\sigma,t)$ to be {\it any} extension of $U$ away from the boundary
such that
\be
U(1,\sigma,t) = U(\sigma,t) ~,~~~ U(0,\sigma,t) = 1
\ee
The WZ action with the proper normalization to reproduce (\ref{chiKM})
is
\be
S_{WZ} = \frac{\hbar^2 \rho_o}{6} \int ds dt \frac{d\sigma}{\sqrt{\theta_o}}
\tr \left\{ U^{-1} \partial_s U [ U^{-1} \LL U , U^{-1} \partial_t U ]
\right\}
\label{SWZ}
\ee

The above WZ action can be written in a more suggestive way by choosing
the $D$-dimensional manifold of integration to be the bulk of the droplet
itself, identifying $s$ with $\phi^D$ and allowing the integration measure
$1/\sqrt{\theta}$ to vary accordingly in the bulk of the droplet. In doing
that, we have to take into account the following:

$\bullet$ $\LL = ( \thRao - \thbao \partial_\beta r_o ) \partial_\alpha$
is defined on the boundary and must be appropriately extended in the bulk

$\bullet$ Unlike $s$, $\phi^D$ is not constant on the boundary, but rather
$\phi^D = r_o (\sigma)$

$\bullet$ The integrand should be a closed form so that its variation
reproduce the same boundary term as (\ref{SWZ})

These actually combine to give the WZ bulk action a simple geometrical
form as the integral of a $D+1$-form, in terms of the canonical two-form
$\omega = \half \omega_{\mu \nu} d\phi^\mu d\phi^\nu$ and the exterior
derivative
$d = dt \partial_t + d\phi^\mu \partial_\mu$. In anticommuting form notation:
\be
S_{WZ} = \frac{\hbar^2 \rho_o}{(k-1)!} \int_D \frac{1}{3}
\omega^{k -1} \tr (U^{-1} dU)^3 ~,~~~ k=\frac{D}{2}
\label{KWZ}
\ee
This has the form of a K\"ahler-Wess-Zumino term on the bulk of the droplet
with $\omega$ playing the role of the K\"ahler structure \cite{NaSc}.
It is obviously a closed form, since both $\omega$ and $\tr (U^{-1}
dU)^3$ are closed ($d\omega=0$ is equivalent to the Jacobi identity for
$\theta$). Its variation will give the boundary integral
\be
\delta S_{WZ} = \frac{\hbar^2 \rho_o}{(k-1)!} \int_{D-1}
\omega_o^{k -1} \tr \left[ (U^{-1} dU)^2 U^{-1} \delta U \right]
\ee
To see that this is what we want, note that $\omega^k$ is a top form
on the phase space and thus
\beqs
\omega^k &=& k! \sqrt{\det\omega} \, d\phi^1 \cdots d\phi^D =
\frac{k!}{(2k)!} \sqrt{\det\omega} \, \epsilon_{\mu_1 \dots \mu_D}
d\phi^{\mu_1} \cdots d\phi^{\mu_D}\cr
\omega^{k-1} &=& \frac{(k-1)!}{2(2k-2)!}
\sqrt{\det\omega} \, \epsilon_{\mu_1 \dots \mu_D} \theta^{\mu_1 \mu_2}
d\phi^{\mu_3} \cdots d\phi^{\mu_D}
\label{omegas}
\eeqs
($\sqrt{\det\omega}$ is really the Pfaffian of the antisymmetric matrix
$\omega_{\mu\nu}$).
Restriction of the form $\omega^{k-1}$ on the boundary will produce
a factor of $\sqrt{\det\omega_o} = 1/\sqrt{\theta_o}$ and will induce the
substitutions
\be
d\phi^D = d\sigma^\alpha \partial_\alpha r_o ~,~~~
d U = dt \partial_t U + d\sigma^\alpha \partial_\alpha U
\ee
So $(U^{-1} dU)^2 = dt d\sigma^\alpha [ U^{-1} \partial_t U , U^{-1}
\partial_\alpha U ]$. Taking into account the combinatorics,
the terms in (\ref{omegas}) with $\mu_1 =D$ or $\mu_2 =D$ reproduce the
term $\theta_o^{D\alpha}$ in $\LL$, while the terms with any of the remaining
$\mu$s equal to $D$ reproduce the term $\theta_o^{\beta\alpha}$ in $\LL$.

Finally, we may recast the full action (\ref{S}) above in a more familiar
form by renaming $\hV_o = -A_0$ and defining the gauged time derivative
\be
D_0 U = \partial_t U - i[ A_0 , U ] = \partial_t U + i [\hV_o , U ]
\ee
The action becomes
\be
S = S_{WZ} -\frac{\hbar^2 \rho_o}{2} \int \frac{dt d\sigma}{\sqrt{\theta_o}} 
\, \tr \left[
U^{-1} (D_0 U - u_o \LL U) U^{-1} \LL U
+ i ( A_0 U^{-1} + U^{-1} A_0 ) \LL U \right]
\label{Sg}
\ee
This has the form of a gauged WZW model.
The last term is the extra term needed in the action to absorb the anomaly
of the Wess-Zumino term under gauge transformations $U \to W^{-1} U W$.
We have recovered the action of Karabali and Nair for a temporal gauge
field, generalized to an arbitrary phase space droplet and with a
$\sigma$-dependent potential gradient $u_o$.

\subsubsection{Comments on the dynamics of the model}

The expression for $\chi=-i U^{-1} \LL U$ implies that the deformation of
the droplet is generated by the unitary field $U$, in analogy with the
spinless case. Specifically, assume a small canonical transformation of the
coordinates $\phi,\xi$ induced by an order-$\hbar$ generating function
$\hbar \Phi (\sigma,\xi)$:
\be
r_o \to r_o + \hbar \thRao \partial_\alpha \Phi ~,~~~
\sigma^\alpha \to \sigma^\alpha + \hbar \thabo \partial_\beta \Phi
~,~~~ \xi^i \to \xi^i + \hbar \thij \partial_j \Phi
\ee
The deformations of $\phi^D = r_o$ and
$\sigma^\alpha$ are of order $\hbar$.
The deformation of $\xi^i$, however, is of order $\hbar^0$ (since
$\thij$ is of order $\hbar^{-1}$) and 
of the same order as $\xi^i$; it cannot be written in the above
infinitesimal form. Instead, we must write the analog of a finite canonical
transformation on the spin space, which is a unitary transformation.

The dynamical meaning of the above WZW structure is as follows: The operator
${\cal L} = u_o \LL$ generates classical motion on the manifold $\phi^D = r_o$
under the spinless part of the single-particle hamiltonian $V_0$.
(The variable $\tau$ represents time of flight along the classical path.)
In the absence of the spin-dependent (matrix) part $\hV$, particles
would simply move along the flow of ${\cal L}$, and so would the
surface of the droplet. Such motion also rescales distances away
from the droplet. The rescaled deformation of the surface $u_o \chi$,
then, would evolve as a co-moving matrix: $u_o {\dot \chi} = {\cal L}
(u_o \chi)$. The trace (scalar) part of $\chi$ represents
total particle number (charge) fluctuations, while its traceless part
represents spin fluctuations of the droplet. The scalar and traceless
parts actually decouple, signaling spin-charge separation in this limit.

In the absence of $\hV$ the motion of the various matrix components of $\chi$
would decouple and could be described as a collection of independent abelian
chiral models. The presence of the spin part $\hV$, however, causes a
`Pauli rotation' of the coordinates $\xi^i$ and thus couples the matrix
components inducing an extra unitary transformation of the matrix $\chi$
that can be understood as gauge (spin) rotation. The gauged WZW action is the
proper dynamical setting for describing such motion.

The appearance of gauge structure in the problem is somewhat surprising,
since we have not introduced gauge fields or considered gauge
transformations. In the next section we shall complete the picture
by doing that.

\section{Introducing gauge fields}

In the analysis so far we have described spin in terms of an internal
compact phase space of the particles. Its canonical structure decoupled
from the one of the kinematical phase space ($\theta^{\alpha i} = 0$)
and any nontrivial spin dynamics arose out of the hamiltonian.

We may further couple spin and kinematical degrees of freedom by
introducing nonzero phase space structure constants between the two
spaces. As we shall demonstrate, this amounts to introducing nonabelian
gauge fields and endows the dynamics with a nonabelian gauge symmetry.
For other examples of introducing gauge degrees of freedom in the
canonical description of particles or fluids see \cite{gaugecan},\cite{JNPP}.

\subsection{Coupling the phase spaces}

The most convenient setting for analyzing the situation is in terms of
the canonical one-form formulation of the phase space. We give below
the relevant facts for our purpose.

We consider a phase space $x^\mu$ endowed with a canonical one-form
${\cal A} = {\cal A}_\mu dx^\mu$ and a hamiltonian $V$. (In our case,
$x^\mu$ will comprise both $\phi^\mu$ and $\xi^i$). $\cal A$ and $V$
could be time-dependent. The phase space action and lagrangian are
\be
L = {\cal A}_\mu {\dot x}^\mu - V ~,~~~ dS = Ldt = A_\mu dx^\mu - V dt
\ee
which leads to the canonical two-form $\omega = d\cal A$ inverse to
the Poisson structure $\theta$:
\be
\omega_{\mu \nu} = \partial_\mu {\cal A}_\nu - \partial_\nu {\cal A}_\mu
~,~~~ \omega_{\mu \nu} \theta^{\nu \rho} = \delta_\mu^\rho
\ee

The above action has the standard phase space invariances.
The first is generated by adding to the lagrangian the total
time derivative of an infinitesimal phase space function $\Phi (x;t)$
\be
\delta_\Phi L = {\dot \Phi}
\ee
which amounts to the abelian gauge transformation
\be
\delta_\Phi {\cal A}_\mu = \partial_\mu \Phi
~,~~~ \delta_\Phi V = - \partial_t \Phi
\ee
leaving the canonical two-form $\omega$ and Poisson structure $\theta$
invariant. The other is general coordinate invariance, generated by
arbitrary infinitesimal coordinate redefinitions
\be
\delta_\epsilon x^\mu = -\epsilon^\mu (x;t)
\label{epsilon}
\ee
which is compensated by the transformation
\be
\delta_\epsilon {\cal A}_\mu = \partial_\nu {\cal A}_\mu \,
\epsilon^\nu + {\cal A}_\nu \,
\partial_\mu \epsilon^\nu ~,~~~ \delta_\epsilon V = \partial_\mu V
\epsilon^\mu - {\cal A}_\mu \partial_t \epsilon^\mu
\ee
(The minus sign in (\ref{epsilon}) is put to stress the fact that
this is a ``passive'' transformation of coordinates.)
The above can be rewritten as
\be
\delta_\epsilon {\cal A}_\mu = -\omega_{\mu \nu} \epsilon^\nu
+ \partial_\mu ({\cal A}_\nu \epsilon^\nu )
~,~~~ \delta_\epsilon V = (\partial_t {\cal A}_\mu + \partial_\mu
V) \epsilon^\mu - \partial_t ({\cal A}_\mu \epsilon^\mu )
\label{cova}
\ee
involving canonically invariant quantities and an abelian gauge
transformation generated by $\Phi_\epsilon = {\cal A}_\nu \epsilon^\nu$.

Canonical transformations are a special case of coordinate transformations
leaving the Poisson structure invariant.
Choosing as coordinate deformation parameters
\be
\epsilon_c^\mu = \{ x^\mu , \Phi \} = \theta^{\mu \nu} \partial_\nu \Phi
\ee
we get for the change in ${\cal A}_\mu$ and $V$
\be
\delta_{\epsilon_c} {\cal A}_\mu = \partial_\mu ( -\Phi
+ \theta^{\nu \rho} {\cal A}_\nu \partial_\rho \Phi )
~,~~~ \delta_{\epsilon_c} V = \{ V , \Phi \} 
+ \partial_t {\cal A}_\mu \theta^{\mu \nu} \partial_\nu \Phi
\ee
This corresponds to an abelian gauge transformation on the ${\cal A}_\mu$
that leaves $\theta^{\mu \nu}$ invariant. For time-independent ${\cal A}_\nu$,
$V$ transforms by a canonical transformation, while a time
dependence in ${\cal A}_\mu$ contributes an extra correction.

We specialize now to the phase space of interest ($\phi^\mu , \xi^i$).
The original lagrangian is
\be
L = {\cal A}_\mu (\phi) {\dot \phi}^\mu + {\cal A}_i (\xi) {\dot \xi}^i - V
\ee
The Poisson structure decouples the spin and kinematical phase
spaces, which reflects to the fact that ${\cal A}_\alpha (\phi)$ and
${\cal A}_i (\xi)$ depend only on their corresponding phase space variables,
ensuring $\partial_i {\cal A}_\alpha = \partial_\alpha {\cal A}_i = 0$.

We shall couple $\xi$ and $\phi$ by relaxing the above condition. In doing
so we do not want to distort the structure of the internal phase space.
Its volume, as well as the area of all noncontractible two-submanifolds,
must remain fixed to appropriate integers for a consistent quantization
(cf. to monopole quantization for $S^2$ and $K$-eigenvalue quantization
for the grassmanian case of section 3.1). It should also stay a homogeneous
space to allow for a linear Poisson algebra in terms of appropriate
spin generators. We shall, therefore, keep its one-form ${\cal A}_i (\xi)$ 
the same as above and independent of $\phi^i$ and shall write it
$\hbar {\bar A}_i$ to explicitly indicate the fact that it is of order
$\hbar$.

We will however allow ${\cal A}_\alpha$ to depend on $\xi$.
The new one-form will consist of the old one, denoted by
${\bar A}_\alpha$, plus a $\xi$-dependent order-$\hbar$ perturbation
${\hbar A}_i$. We also write the hamiltonian in the form ${\bar V} + \hbar \hV
= {\bar V} - \hbar A_0$, taking a hint from the previous section in
renaming $\hV$ to $-A_0$. Further, we will allow $A_\alpha$ and $A_0$
to be time-dependent.  Altogether the lagrangian becomes
\be
L = {\bar A}_\alpha (\phi) \, {\dot \phi}^\alpha +\hbar{\bar A}_i (\xi)
{\dot \xi}^i -{\bar V} (\phi) + \hbar
A_\alpha (\phi,\xi,t) \, {\dot \phi}^\alpha + \hbar A_0 (\phi,\xi,t)
\label{Acouple}
\ee
In terms of scales, $\omega_{ij} = \hbar(\partial_i {\bar A}_j -
\partial_j {\bar A}_i )$
is of order $\hbar$ and $\theta^{ij} = (\omega^{-1} )^{ij}$ is of
order $1/\hbar$ as it should. 

\subsection{Gauge transformations}

We come now to the issue of gauge transformations in the spin space
of our phase space structure (\ref{Acouple}). Interpreting $\xi$ as
spin variables, we understand that gauge transformations should amount to
local rotations of the $\xi$-coordinates in their phase space; that is,
canonical transformations in the $\xi$ space that depend on the kinematical
phase space coordinates. These will be generated by an order-$\hbar$
function $\hbar \Phi (\phi,\xi,t)$
\be
\delta \xi^i = -\hbar \{ \xi^i , \Phi \}_\xi 
= -\hbar \theta^{ij} \partial_j \Phi ~,~~~
\delta \phi^\alpha = 0
\ee
Note that the above is {\it not} a canonical transformation on the full
space, since the kinematical coordinates $\phi$ are not transformed
and the Poisson bracket in the transformation of $\xi$ is restricted.
The canonical one-form and hamiltonian are transformed according to
(\ref{cova}), which implies for $A_\alpha$, $A_0$ and ${\tilde A}_i$:
\beqs
\delta A_\alpha &=& \hbar \,\partial_i A_\alpha \,\theta^{ij} \partial_j \Phi
+\hbar \partial_\alpha (\theta^{ij} A_i \partial_j \Phi ) \cr
\delta A_0 &=& \hbar \,\partial_i A_0 \,\theta^{ij} \partial_j \Phi
+\hbar \partial_0 (\theta^{ij} A_i \partial_j \Phi ) \cr
\delta {\bar A}_i &=& -\partial_i \Phi 
+ \hbar \partial_i ( \theta^{ij} {\bar A}_i \partial_j \Phi )
\eeqs
We observe that $A_\alpha$ and $A_0$ transform as the space and time
components of a one-form. Calling $x^0 =t$ and using middle-greek letters for
spacetime indices, $\mu,\nu = 0,1,\dots D$ (not to be confused with
$\mu,\nu = 1,2,\dots D$ used in early sections), we can write the above as
\beqs
\delta A_\mu &=& \hbar \{ A_\mu , \Phi \}_\xi
+\partial_\mu (\hbar \theta^{ij} {\bar A}_i \partial_j \Phi ) \cr
\delta {\bar A}_i &=& \partial_i ( -\Phi 
+ \hbar \theta^{ij} {\bar A}_i \partial_j \Phi )
\eeqs
Under the above transformation, the spin one-form transforms away from
its reference value ${\bar A}_i$ by a total derivative.
We may restore it to its original form by adding to the lagrangian the
total derivative of $\Phi - \hbar \thij {\bar A}_i \partial_j \Phi$.
This further transforms $A_\mu$ to
\be
\delta_g A_\mu = \partial_\mu \Phi + \hbar \{ A_\mu , \Phi \}_\xi
\ee
This has exactly the form of a nonabelian gauge transformation. Indeed,
upon quantization of the spin space, $\xi$-dependent quantities such as
$\Phi$ and $A_\alpha$ become matrices and Poisson brackets become
($1/i\hbar$) commutators. So the above transform becomes
\beqs
\delta_g A_\mu &=& \partial_\mu \Phi - i [ A_\mu , \Phi ] \cr
\delta \xi^i &=& i [ \xi^i , \Phi ]
\eeqs
which corresponds to the transformation of a covariant derivative
$D_\mu = \partial_\mu -i A_\mu$ and an anti-covariant spin matrix
$\xi^i$ under an infinitesimal unitary rotation $U = 1+i\Phi$: 
\beqs
D_\mu &\to & U D_\mu U^{-1} \cr
\xi^i &\to & U^{-1} \xi^i U
\eeqs
The $A_\mu$, thus, can be rightfully considered as nonabelian gauge
fields on the phase space. Gauge transformations are local
rotations of the spin coordinates, which is a passive transformation,
reflected in the anti-covariant nature of $\xi^i$.
We recover the dynamics of a single spinning (in fact,
colored) particle interacting with a nonabelian gauge field in phase
space.

The group of the nonabelian gauge transformations is determined by the
realization of the spin phase space. Gauge invariance derives from
canonical spin transformations and therefore inherits the full $SU(n)$
symmetry group of the spin Hilbert space. Its realization, however,
may be restricted by the physics of the problem (cf. the discussion
of section 3.2.2).

As an example, in the realization in terms of a spin-$\frac{n-1}{2}$
representation of $SU(2)$, as argued in section 3.2.2, only operators
linear in the spin variables $S^i = \hbar \xi^i$ are natural and thus
the expression for $A_\mu$ will be restricted to
\be
A_\mu (\phi,\xi,t) = A_\mu^i (\phi,t) S^i ~,~~~ i=1,2,3
\ee
A general $SU(n)$ transformation will take the above $A_\mu$ away from
this form. Only unitary transformations in the $SU(2)$ subgroup
transforming linearly the $\xi^i$ will be proper gauge transformations,
the field $A_\mu$ being in the spin-$\frac{n-1}{2}$ representation of
the group. The grassmanian representation with a Kirillov form
$K = diag(n_1 , \dots n_M )$ and a corresponding linear restriction
for the form of $A_\mu$ would produce an $SU(M)$ gauge group in the
representation corresponding to Young tableau with $n_i$ blocks in
row $i$.

\subsection{Equations of motion}

The above interpretation can be further justified
by looking at the single-particle equations of motion in the presence
of the extra coupling between $\phi$ and $\xi$ due to the (time-dependent)
$A_\alpha$ and $A_0$. These can be obtained by varying the action with
an arbitrary $\epsilon^\mu$ and setting the variation to zero. Using
(\ref{cova}) this implies
\be
\omega_{\mu \nu} {\dot x}^\nu = 0
\label{covem}
\ee
where $x^\mu$ stands for $\{ x^0 , \phi^\alpha , \xi^i \}$.
(In principle, since we do not vary $x^0 =t$, we only obtain the above
equations for $\mu \neq 0$. The $\mu=0$ equation, however, holds true
as a corollary of the remaining equations, due to the identity
$\omega_{\mu \nu} {\dot x}^\mu {\dot x}^\nu = 0$, so we obtain the
full covariant set of equations.)
These can also be written as 
\be
{\dot x}^\alpha = \theta^{\alpha \beta} ( \partial_\beta V + \partial_0
A_\beta ) = \{ x^\alpha , V \}_{sp} + \theta^{\alpha \beta} \partial_0 A_\beta
~~~(\alpha,\beta \neq 0)
\ee
which, for time-independent $A_\alpha$, reduce to the usual canonical
equations of motion.

Applying the above equation (\ref{covem}) to the case of the
lagrangian (\ref{Acouple}) for $\phi^\alpha , \xi^i$ we obtain
\beqs
\left[ {\bar \omega}_{\alpha \beta} + \hbar (\partial_\alpha A_\beta
- \partial_\beta A_\alpha ) \right] {\dot \phi}^\beta &=&
\partial_\alpha {\bar V} +\hbar (\partial_0 A_\alpha -\partial_\alpha A_0
+\hbar \partial_i A_\alpha {\dot \xi}^i ) \cr
{\bar \omega}_{ij} {\dot \xi}^j + \hbar ( \partial_i A_0 +
\partial_i A_\alpha {\dot \phi}^\alpha ) &=& 0
\eeqs
where ${\bar \omega} = d {\bar {\cal A}}$ is the reference 
(uncoupled) canonical two-form.
The above equations can be combined and rewritten as
\beqs
&&({\bar \omega}_{\alpha \beta} + \hbar F_{\alpha \beta} ) {\dot \phi}^\beta
= \partial_\alpha {\bar V} + \hbar F_{0\alpha} \cr
&&{\dot \xi}^i - \hbar \{ A_0 + {\dot \phi}^\alpha A_\alpha , \xi^i \}_\xi = 0
\eeqs
where
\be
F_{\mu \nu} = \partial_\mu A_\nu - \partial_\nu A_\mu + \hbar
\{ A_\mu , A_\nu \}_\xi
\ee
is the nonabelian field strength of the gauge field $A_\mu$. These
equations have the structure of the equations of motion of a particle
with nonabelian degrees of freedom and only involve gauge covariant
quantities: the first is the standard `minimal' coupling of the
particle's coordinates to a field strength coupled to its (nonabelian)
charge (with a scalar potential $\bar V$ and an electric field
$F_{0\alpha}$), while the second is the covariant parallel transport of the
spin over the particle's phase-space-time trajectory.
Due to the anti-covariant (passive) nature of $\xi$,
its equation of motion involves covariant derivatives with the opposite
sign for $A_\mu$. Observables of the form
\be
Q = \int d\xi \, Q_i \xi^i \to \tr (Q_i \xi^i )
\ee
are gauge invariant, while $Q_i$ transform covariantly.

\subsection{Gauged droplet dynamics}

The generalization of the droplet construction for the gauged phase space
considered above is straightforward. The Poisson structure is, now,
time-dependent, involving the gauge fields $A_\alpha$, but the counting
of states and fermion exclusion principle that led to constant-density
droplets remain the same. The construction of the boundary field
Poisson brackets, hamiltonian and equation of motion for the classical
theory is as in section 3.2 with the generalized form of $\theta_b^{
\alpha \beta}$, involving gauge fields, appearing in the formulae.

The quantization of the spin space in this case presents some new
ordering ambiguities, since we cannot any more assume that $\theta^{
\mu \nu}$ is independent of $\phi^D$ and $R$. The proper ordering of
the full nonlinear matrix Poisson brackets for $R^{ab}$ will be
partly determined by the requirement that they satisfy the Jacobi
identity.

To leading-order in $\hbar$, however, there are no ambiguities.
All nonlinear terms in the Poisson structure that would require ordering
are of higher order and can be ignored. The leading terms reproduce a
fully gauged Kac-Moody algebra, as we shall demonstrate.

The canonical two-form as derived from (\ref{Acouple}), denoted
${\tilde \omega}_{\mu \nu}$,
consists of a leading part $\omega_{\mu\nu}$ and an order-$\hbar$ part:
\beqs
{\tilde \omega}_{\alpha \beta} &=& \omega_{\alpha \beta} + 
\hbar ( \partial_\alpha A_\beta - \partial_\beta A_\alpha ) \cr
{\tilde \omega}_{i\alpha} &=& \hbar \partial_i A_\alpha \cr
{\tilde \omega}_{ij} &=& \omega_{ij} = \hbar
(\partial_i {\bar A}_j - \partial_j {\bar A}_i )
\eeqs
The expansion of ${\tilde \theta} = {\tilde \omega}^{-1}$ in $\hbar$
is complicated by the fact that, to leading order, $\omega_{ij}$
vanishes and so the $\hbar^0$ part of $\omega_{\mu \nu}$ is
singular. To overcome this, we temporarily change the scale of the
spin coordinates by incorporating a factor of $\sqrt \hbar$ in each
$\xi^i$, which has the effect:
\be
\omega_{i\alpha} \to \hbar^{-\half} \omega_{i\alpha} ~,~~~
\omega_{ij} \to \hbar^{-1} \omega_{ij}
\ee
The rescaled $\omega$ becomes:
\beqs
{\tilde \omega}_{\alpha \beta} &=& {\bar \omega}_{\alpha \beta} + 
\hbar ( \partial_\alpha A_\beta - \partial_\beta A_\alpha ) \cr
{\tilde \omega}_{i\alpha} &=& \hbar^\half \partial_i A_\alpha \cr
{\tilde \omega}_{ij} &=& {\bar \omega}_{ij} = \partial_i {\bar A}_j -
\partial_j {\bar A}_i 
\eeqs
This is an order $\hbar^\half$ perturbation $\delta \omega$
over a nonsingular form ${\bar \omega}_{\mu \nu}$. We can calculate
the inverse in the standard expansion, 
\be
{\tilde \theta} = {\bar \theta} - {\bar \theta} \delta \omega {\bar \theta} 
+ {\bar \theta} \delta \omega {\bar \theta} \delta \omega {\bar \theta}
+ \dots
\ee
(${\bar \theta} = {\bar \omega}^{-1}$).  The result to order $\hbar$ is:
\beqs
{\tilde \theta}^{\alpha \beta} &=& {\bar \theta}^{\alpha \beta} -
\hbar {\bar \theta}^{\alpha \gamma}
F_{\gamma \delta} {\bar \theta}^{\delta \beta} \cr
{\tilde \theta}^{\alpha i} &=& \hbar^\half {\bar \theta}^{\alpha \beta}
{\bar \theta}^{ji} \partial_j A_\beta \cr
{\tilde \theta}^{ij} &=& {\bar \theta}^{ij} + \hbar {\bar \theta}^{ik}
{\bar \theta}^{jl} {\bar \theta}^{\alpha \beta} \partial_k A_\alpha
\partial_l A_\beta
\eeqs
We see that we now have a nonvanishing $\theta^{\alpha i}$.
Finally, we may restore the original scale of the spin coordinates,
which amounts to ${\tilde \theta}^{\alpha i } \to \hbar^{-\half} 
{\tilde \theta}^{\alpha i}$, ${\tilde \theta}^{ij} \to \hbar^{-1}
{\tilde \theta}^{ij}$. We also revert to the original
spin Poisson structure, $\theta^{\alpha \beta} = {\bar \theta}^{\alpha
\beta}$, $\theta^{ij} = \hbar^{-1} {\bar \theta}^{ij}$.
The final result is
\beqs
{\tilde \theta}^{\alpha \beta} &=& \theta^{\alpha \beta} -
\hbar \theta^{\alpha \gamma} F_{\gamma \delta} \theta^{\delta \beta} \cr
{\tilde \theta}^{\alpha i} &=& \hbar \theta^{\alpha \beta}
\theta^{ji} \partial_j A_\beta \cr
{\tilde \theta}^{ij} &=& \theta^{ij} + \hbar^2 \theta^{ik}
\theta^{jl} \theta^{\alpha \beta} \partial_k A_\alpha
\partial_l A_\beta
\label{thetaperturb}
\eeqs
Similarly, the determinant $\det \theta = (\det \omega)^{-1}$
will receive corrections according to
\be
\det \omega = \det {\bar \omega} \left[ 1 + \tr ({\bar \theta} \delta
\omega) - \half \tr ({\bar \theta} \delta \omega)^2 + \half [ \tr
({\bar \theta} \delta \omega )]^2 + \dots \right]
\ee
and these will be of higher order in $\hbar$.

We may now use the new expressions (\ref{thetaperturb}) in the Poisson brackets
for the boundary field (\ref{PRCS}). The new terms ${\tilde \theta}^{\alpha i}$
appear with derivatives acting on $R$ or $\delta$. Such terms, acting on
a function $g$ (=$R$ or $\delta$) create new
terms in the Poisson brackets of the form 
\be
\theta^{\alpha i} \partial_i g = \hbar \theta^{\alpha \beta} \theta^{ji}
\partial_j A_\beta \partial_i g = \hbar \theta^{\alpha \beta}
\{ A_\beta , g \}_\xi 
\ee
Upon quantization of the spin space, $\hbar \{ A_\beta , g \} \to
-i [ A_\beta , g ]$ and the above terms become commutators. Combined
with the corresponding term $\theta^{\alpha \beta} \partial_\beta$
they give
\be
\theta^{\alpha \beta} \partial_\beta g + \theta^{\alpha i} \partial_i g
= \theta^{\alpha \beta} (\partial_\beta g -i [A_\beta , g] )
= \theta^{\alpha \beta} D_\beta g
\ee
Their net effect is to gauge all the derivatives appearing in the
Poisson brackets. This is the leading change in $\hbar$. Other terms
will produce higher order effects. For instance, the new term in
${\tilde \theta}^{ij}$ will produce the term
\be
\theta^{\alpha \beta} [R , A_\alpha ] [A_\beta , \delta ]
\ee
Although this does not involve explicit factors of $\hbar$,
upon putting $R = r_o + \hbar \chi$ the contribution of the scalar
leading term $R_o$ vanishes and the above term is of order $\hbar$.

Altogether and obtain
\be
\{ \chi^A (\sigma_1 ) , \chi^B (\sigma_2 ) \} =
\frac{\sqrt{\theta_{o+}}}{\hbar^2\rho_o} \Bigl[
(\theta_{o+}^{0\alpha} - \theta_{o+}^{\beta\alpha} \,\partial_\beta r_o )
D_\alpha^{AB} \delta(\sigma_- )
+ f^{ABC} \chi^C \, \delta (\sigma_- ) \Bigr]
\label{gauAB}
\ee
where $D_\alpha^{AB}$ is the adjoint expression of the covariant derivative
$D_\alpha$
\be
D_\alpha^{AB} = \delta^{AB} \partial_\alpha - f^{ABC} A_\alpha^C
\ee
and $D_\beta r_o = \partial_\beta r_o$ since $r_o$ is a scalar.
Similarly, the hamiltonian obtains as
\be
H = \hbar^2 \rho_o \int \frac{d\sigma}{\sqrt{\theta_o}} \, \tr \left(
\half u_o \chi^2 - A_0 \chi \right) 
\ee
We may define as before a derivative along the direction of classical
motion on the boundary and the corresponding covariant version:
\be
\DD = (\thRao - \thbao \partial_\beta r_o ) D_\alpha
\ee
In terms of $\DD$ the equation of motion becomes
\be
D_0 \chi - \DD ( u_o \chi) = F_{0\tau}
\label{gaugEOM}
\ee
where
\be
F_{0\tau} = [ D_0 , \DD ]
\ee
Similarly, the Poisson brackets of $\chi$ become
\be
\{ \chi^A (\sigma_1 ) , \chi^B (\sigma_2 ) \} =
\frac{\sqrt{\theta_{o+}}}{\hbar^2\rho_o} \Bigl[
\DD^{AB} (\sigma_- ) \delta(\sigma_- ) 
+ f^{ABC} \chi^C \, \delta (\sigma_- ) \Bigr]
\label{gauKM}
\ee
We obtain a gauged Kac-Moody algebra and corresponding equation of motion
for the chiral current $\chi = J_+$. As before, this structure derives
from the action
of a fully gauged Wess-Zumino-Witten model in the space defined by 
classical motion trajectories and time, integrated over the remaining
phase space variables. In terms of a unitary field $U$ we have
\be
\chi = -i U^{-1} \DD U
\ee
and the action is
\beqs
S &=&  \hbar^2 \rho_o \int \frac{dtd\sigma}{\sqrt{\theta_o}} \,
\tr \left[ -\half U^{-1} (D_0  - u_o  \DD )U \, U^{-1} \DD U \right] \cr
&+& \frac{\hbar^2 \rho_o}{(k-1)!} \int_{D-1} \omega_o^{k-1} \tr 
(A U^{-1} dU + A dU U^{-1} + A U^{-1} AU) \cr
&+& \frac{\hbar^2 \rho_o}{(k-1)!} 
\int_D \frac{1}{3} \omega^{k -1} \tr (U^{-1} dU)^3 
\label{GWZW}
\eeqs
with $D=2k$. The first term is the gauged kinetic term on the boundary.
The last term is the standard Wess-Zumino term; it 
does not involve gauge fields and is obtained by integrating the
Wess-Zumino form over the bulk of the droplet with an appropriately
extended unitary field $U$ as in section 3.3. The middle term is
defined on the boundary of the droplet and involves the gauge fields
$A_0$ and $A_\tau$; it is needed to absorb the gauge non-invariance of
the Wess-Zumino term and contributes the term $F_{0\tau}$
in the equation of motion for $\chi$ (\ref{gaugEOM}).

Overall, we have recovered the action of \cite{KaNa} for a fully general
gauge field, generalized to an arbitrary phase space droplet and with a
$\sigma$-dependent potential gradient $u_o$.

\section{Conclusions and discussion}

We have presented an analysis of the phase space dynamics of droplets
representing fermions with internal degrees of freedom in an arbitrary
phase space and derived their hamiltonian and canonical structure.

To leading order in $\hbar$ we recovered the WZW chiral action
of edge excitations.
In the nonlinear theory we do not have an explicit form for the action.
This is not crucial, since we have derived the complete hamiltonian dynamics,
but it remains an issue for further investigation, especially if we are
interested in applying path-integral or effective field theory techniques.

The nature of the obtained theories is halfway between classical and
quantum: spin is quantized and gives rise to a matrix structure, while
phase space coordinates are still treated classically. As such, it is
reminiscent of the matrix formulation of the quantum Hall effect
\cite{matHall}. The exact correspondence between the two formulations,
if any, should be further examined.

Finally, the theories derived in this paper represent a nonabelian phase
space bosonization of the fermionic systems they describe. Just as in the
abelian case, however, this bosonization fails quantum mechanically in
dimensions higher than $D=2$. The main problems are, first, that this theory
overestimates the degrees of freedom of the system, due to the infinity
of excitations normal to the direction of propagation and, second, that
the theory is essentially local in phase space and thus does not take
into account processes where fermions would undergo transitions to faraway
phase space states. This is an issue that will be treated in an upcoming
publication.

\vskip 0.1in

{\it \underline {Acknowledgements}:} I would like to thank
D.~Karabali and V.P.~Nair for useful comments on the manuscript.
This research was supported in part by the
National Science Foundation under grant PHY-0353301 and by the CUNY 
Research Foundation under grant PSC-CUNY-66565-0035.

\end{document}